\documentclass[conference]{IEEEtran}

\usepackage{tikz,pgfplots}
\usepackage{fancyhdr,graphicx,amsmath}
\usepackage{amsfonts} 
\usepackage{float}
\restylefloat{table}
\usepackage{chngpage}

\usepackage{graphics}
\usepackage{balance} 
\usepackage{booktabs}
\usepackage{makecell}
\usepackage{hyperref}
\usepackage{caption}
\usepackage{url} 

\graphicspath{ {picture/} } 
\usepackage{tabularx} 
\usepackage{times}
\usepackage{array,multirow}
\usepackage{makecell}
\usepackage{tabularx, multirow, makecell, rotating} 
\usepackage{xcolor}

\newcommand{\rev}[1]{{{#1}}}
\newcommand{\eat}[1]{}

\usepackage{amsmath}
\usepackage{color}
\usepackage{colortbl}
\usepackage{epsfig}
\usepackage{xspace}
\usepackage{wrapfig}
\usepackage{graphicx}
\usepackage{subcaption}
\usepackage{enumerate}
\usepackage[color,matrix,arrow,all]{xy}
\usepackage{booktabs}
\usepackage{balance}
\usepackage{stmaryrd}
\usepackage{hhline}
\usepackage{paralist}
\usepackage{setspace}
\usepackage{comment}
\usepackage[ruled,vlined,linesnumbered,shortend]{algorithm2e}

\usepackage{epsfig}
\usepackage{multirow}
\usepackage{url}

\definecolor{darkgreen}{rgb}{0.2,0.7,0.2}

\newcommand{\wvec}{\textsc{w2vec}\xspace}
\newcommand{\dvec}{\textsc{d2vec}\xspace}
\newcommand{\fast}{\textsc{f-txt}\xspace}
\newcommand{\sBert}{\textsc{s-be}\xspace}
\newcommand{\tapas}{\textsc{tapas$^*$}\xspace}
\newcommand{\lBert}{\textsc{l-be}\xspace}
\newcommand{\bBert}{\textsc{b-be}\xspace}
\newcommand{\dBert}{\textsc{d-be}\xspace}
\newcommand{\rank}{\textsc{rank}\xspace}
\newcommand{\rwWord}{\textsc{\underline{w-rw}}\xspace}
\newcommand{\rwWordEx}{\textsc{\underline{w-rw-Ex}}\xspace}
\newcommand{\rwFast}{\textsc{\underline{f-rw}}\xspace}

\newcommand{\sBertS}{\textsc{s-be$^*$}\xspace}
\newcommand{\lBertS}{\textsc{l-be$^*$}\xspace}
\newcommand{\bBertS}{\textsc{b-be$^*$}\xspace}
\newcommand{\dBertS}{\textsc{d-be$^*$}\xspace}
\newcommand{\rankS}{\textsc{rank$^*$}\xspace}
\newcommand{\ditto}{\textsc{ditto$^*$}\xspace}
\newcommand{\deep}{\textsc{deep-m$^*$}\xspace}

\newlength\mylen
\newcommand\myinput[1]{%
  \settowidth\mylen{\KwIn{}}%
  \setlength\hangindent{\mylen}%
  \hspace*{\mylen}#1\\}

\newcommand{\bi}{\begin{itemize}}
\newcommand{\ei}{\end{itemize}}

\newcommand{\be}{\begin{enumerate}}
\newcommand{\ee}{\end{enumerate}}
\newcommand{\beqn}{\begin{eqnarray*}}
\newcommand{\eeqn}{\end{eqnarray*}}

\newcounter{ccc}

\newcommand{\eop}{\hspace*{\fill}\mbox{$\Box$}\vspace{1ex}}     
\newcounter{example}
\renewcommand{\theexample}{\arabic{example}}
\newenvironment{example}{
        \vspace{1ex}
        \refstepcounter{example}
        {\noindent\bf Example \theexample:}}{
        }
\newcounter{theorem}
\renewcommand{\thetheorem}{\arabic{theorem}}
\newcounter{lemma}
\newcounter{definition}
\renewcommand{\thedefinition}{\arabic{definition}}

\definecolor{shadecolor}{RGB}{200,200,200}

\definecolor{shadecolor1}{RGB}{230,230,230}

\definecolor{shadecolor1}{RGB}{255, 114, 118}

\begin{document}
\title{Unsupervised Matching of Data and Text}

\author{\IEEEauthorblockN{Naser Ahmadi}
\IEEEauthorblockA{EURECOM, France\\
naser.ahmadi@eurecom.fr}
\and
\IEEEauthorblockN{Hansj\"org Sand}
\IEEEauthorblockA{KPMG, Germany\\
hsand@kpmg.com}
\and
\IEEEauthorblockN{Paolo Papotti}
\IEEEauthorblockA{EURECOM, France\\
paolo.papotti@eurecom.fr}}

\maketitle
\begin{abstract}
Entity resolution is a widely studied problem with several proposals to match records across relations. Matching textual content is a widespread task in many applications, such as question answering and search. While recent methods achieve promising results for these two tasks, there is no clear solution for the more general problem of matching textual content and structured data. We introduce a framework that supports this new task in an unsupervised setting for any pair of corpora, being relational tables or text documents. Our method builds a fine-grained graph over the content of the corpora and derives word embeddings to represent the objects to match in a low dimensional space. The learned representation enables effective and efficient matching at different granularity, from relational tuples to text sentences and paragraphs. Our flexible framework can exploit pre-trained resources, but it does not depends on their existence and achieves better quality performance in matching content when the vocabulary is domain specific. We also introduce optimizations in the graph creation process with an ``expand and compress'' approach that first identifies new valid relationships across elements, to improve matching, and then prunes nodes and edges, to reduce the graph size. Experiments on real use cases and public datasets show that our framework produces embeddings that outperform word embeddings and fine-tuned language models both in results' quality and in execution times. 
\end{abstract}

\section{Introduction}
In data integration, matching records referring to the same real world object is an important task, usually referred to as \textit{entity resolution} (\textbf{ER})~\cite{CappuzzoPT20,DITTO21,getoor2013entity}. In other communities, such as in Natural Language Processing (NLP), \textit{text matching} (\textbf{TM}) is also a widespread task in many applications, such as question answering~\cite{RajpurkarJL18} and information retrieval~\cite{guo2019matchzoo}.
However, in many scenarios the borders between the two tasks are not clearly defined. Several datasets have long textual cell values, such as product descriptions. Text documents have structural properties and content organized in hierarchies. Finally,
some applications \textit{match textual content to structured data}, such as relational tuples~\cite{ChenWCZWLZW20,KaragiannisSPT20}. 

The best \textbf{ER} results are obtained by methods exploiting deep learning techniques~\cite{CappuzzoPT20,EbraheemTJOT18,DITTO21,mudgal2018deep}, but \rev{they rely on the presence of multiple attributes (or fields) in the schema}, which is missing in text.
Transformer-based approaches have enabled important improvements in \textbf{TM}~\cite{devlin2018bert,bigbird20}. 
However, transformers are designed to capture the (hidden) relationships in the language. \rev{Methods designed to learn table relationships need a large corpus in the pre-training~\cite{YinNYR20,HerzigNMPE20,DengSL0020} and their pre-trained models do not achieve top quality performance on unseen tables for our task, as we show experimentally.}

Consider the following examples.

\begin{figure}[ht]
  \includegraphics[width=0.51\textwidth]{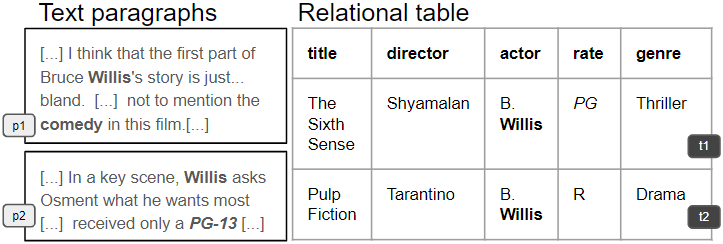}
  \vspace{-2ex} 
  \caption{Text and data: paragraph $p1$ matches tuple $t2$.}
  \label{fig:movies}
\vspace{-4.5ex}
\end{figure}

\begin{example}
\label{ex:1}
\textbf{Text and relational data.} A corpus of product reviews is gathered from the Web.  
A company must link tuples in a relation to these reviews for
a promotional campaign.
However, the product reviews have no identifier (Figure~\ref{fig:movies}). 
\end{example}

\begin{figure}[ht]
\includegraphics[width=0.5\textwidth]{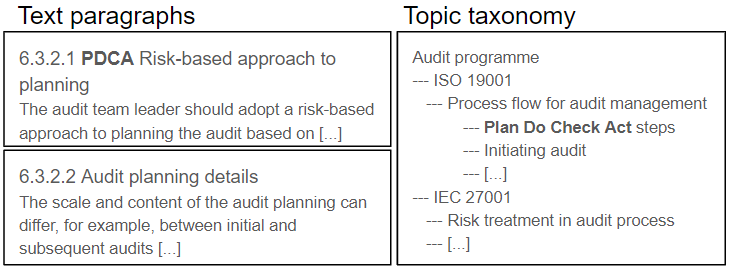}
  \vspace{-3ex}
  \caption{Structured texts: 1$^{st}$ paragraph matches 4$^{th}$ node.}
  \label{fig:kpmg}
\vspace{-2.5ex}
\end{figure}

\begin{example}
\label{ex:2}
\textbf{Structured texts.} An enterprise manual about auditing processes is hard to navigate for the final users. To support search, the paragraphs in the manual must be matched to a large taxonomy of concepts (Figure~\ref{fig:kpmg}). 
\end{example}

\vspace{1ex}
This problem is difficult  
because matching is based on overlapping (or similar) content in the input objects, but this signal can be missing or ambiguous in this new setting. 

{\bf Missing matches.} In Example~\ref{ex:1}, 
movie \textit{Pulp Fiction} is reported as \textit{Drama} in the table but \textit{comedy} is mentioned in the review. This problem can be partially tackled with the pre-trained embedding of \textit{Tarantino}, which models him as a director for both comedies and dramas~\cite{pennington2014glove}. But in Example~\ref{ex:2}, modeling the connection between \textit{PDCA} and its full spelling is crucial to match the paragraph to the right taxonomy node. While pre-trained embeddings can be used to identify synonyms for common words and popular entities, they fail for \textit{domain specific} terms.
\textbf{Challenge 1}: as specific vocabularies are not well modeled by pre-trained resources, we need to learn embeddings \textit{across the heterogeneous corpora} at hand to discover similarities in their content.

\begin{figure*}[ht]
\centering
  \includegraphics[width=0.9\textwidth]{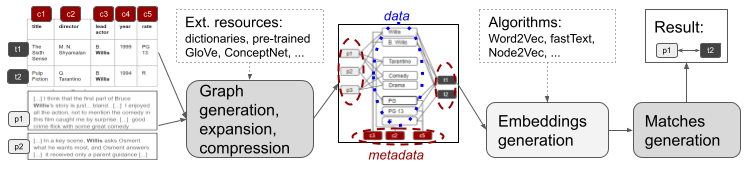}
    \vspace{-1ex}
  \caption{The proposed unsupervised solution: (1) jointly model text and structured data documents in a graph, (ii) produce embeddings for data and metadata nodes (representing texts, taxonomy nodes, tuples), (iii) match metadata nodes.}
  \label{fig:process}
   \vspace{-2ex}
\end{figure*}

{\bf Ambiguous matches.} In Example~\ref{ex:1}, an actor named \textit{Willis} appears in different paragraphs and tuples, but only one tuple in the relation is the correct match in this case. Similarly, the term \textit{audit} appears in most paragraphs and taxonomy nodes in the second example. This suggests the need of a weighting mechanism for combining the matching tokens across two candidate objects. 
\textbf{Challenge 2}: there is the need to learn how to combine matching signals, but the lack of training data rules out solutions based on fine tuning of  pre-trained models.

\vspace{1ex}
Previous methods lack the flexibility to cover such challenges.
Existing approaches handle traditional \textbf{ER} and \textbf{TM}, but they fail in terms of accuracy in the new use cases above. 
What is missing is a unified representation that is at the same time modeling the \textit{relationships in the structured content} (for learning a good representation and identify similarities) and the \textit{importance of every matching word} when comparing heterogeneous objects. The last point highlights the need of learning a comparable representation for sets and sequences of tokens, such as tuples and text documents.

To overcome these issues, we propose a framework for learning representations of data and text that (i) is tailored at the domain at hand with a joint modeling of heterogeneous corpora and (ii) exploits structured information whether available to improve the quality of the generated embeddings and of the matching process. 

Figure~\ref{fig:process} shows our solution. 
First, it represents text documents and tables as nodes and edges in an undirected graph. This graph contains two main types of nodes. \textit{Data} nodes represent tokens (words) in the corpora, either in text paragraphs or in table cells. \textit{Metadata} nodes represent IDs for tuples, attributes and paragraphs. Graph edges represent the relationship between data and metadata, e.g., a tuple/attribute/paragraph contains the token in a data node. 
As our goal is to match metadata nodes, we aim at creating more paths between related nodes and at removing spurious connections. The first goal is achieved in an expansion step that exploits external resources, such as ConceptNet~\cite{speer2017conceptnet}. The second goal is obtained by pruning edges and nodes with graph compression techniques designed for our matching task.
Next, we generate an embedding for every graph node. 
We rely on existing solutions for this step and the algorithm at hand can be replaced as the community makes progress in this task.
Finally, we use the embeddings for the metadata nodes in an unsupervised algorithm to identify the matching ones, such as the paragraph and the tuple in the first example.

The framework enables users to improve the solution according to the requirements and the resources at hand. If relevant external resources exist, such as word dictionaries, they can be plugged in graph construction to merge data nodes. Knowledge graphs and ontologies are plugged in the expansion step to find more relationships across metadata nodes. New embedding generation algorithms can be plugged in the second step to improve the quality of the embeddings.

Our work extend recent solutions designed for the \textbf{ER} task (data to data only) to this new problem~\cite{CappuzzoPT20,abs-2010-07386,DITTO21} \rev{by enabling matching for schema-less text paragraphs and taxonomies}. The module for graph generation is our first contribution as it creates a rich representation which is then reflected in the \textit{domain specific} embeddings for text and structured data metadata nodes, thus tackling the first challenge. 
Our second contribution is the expansion and compression approach that, together with matching generation module, exploits the benefits of embeddings metadata nodes in an \textit{unsupervised solution}, thus tackling the second challenge. \rev{While our work extends~\cite{CappuzzoPT20}, this paper introduces the novel problem of matching text paragraphs to tuples and nodes in taxonomies, it extends the original graph to data sources without a schema, and it handles long free texts with a novel expansion and compression approach.}
Our proposal outperforms state-of-the-art methods by increasing quality performance up to 45\% in absolute terms while taking a fraction of their time in matching. Finally, while we focus on unsupervised applications, any downstream classifier can be trained using the embeddings from our solution.

In the following, Section~\ref{sec:method} describes the proposed 
framework and how we generate and refine the graph at the core of the proposal. Section~\ref{sec:expand} introduces algorithms to expand the original graph with external resources and compress it to keep its size manageable. 
Section~\ref{sec:secondStep} introduces the methods for producing embeddings and matching metadata nodes.
Section~\ref{sec:experiments} evaluates our work with datasets from real applications. 
Section~\ref{sec:related} discusses related work. 
Section~\ref{sec:conclusion} concludes the paper with open challenges and the next steps we plan to take. 

\section{A Graph for Heterogeneous Corpora} \label{sec:method}

In this section, we describe the algorithm for the generation of a graph across heterogeneous corpora.  
We discuss the case with text documents and tables as corpora, but the same algorithm applies for the case with documents (tables) only. 

The input of the graph creation are two corpora.
A \textit{corpus}, based on the task, is a table, some structured text, or simple text. The \textit{document} to match is a tuple, for tables, while the granularity of the text is user-defined and can span from a single sentence to a paragraph. 
These corpora are matched in three possible combinations, or \textit{tasks}: text to structured text matching, text to data matching, or text to text matching. The purpose of each task is to find the top-$k$ closest documents in the second corpus for all the documents in the first corpus. External resources, such as pre-trained embeddings, are not needed to run the pipeline, but 
they can be naturally exploited as we discuss in Section~\ref{subsec:external}.

We jointly represent the document corpora and tables in a graph, from which we then generate embeddings. We first perform some pre-processing steps for every corpus. This includes stop-words removal and stemming on the tokens coming from the texts and the cell values of the tables. We call \textit{terms} these processed values and a term can be composed of one or multiple tokens. For example, ``The Sixth Sense" is a term composed of three tokens. 

We define two types of nodes. \textit{Data nodes} represent the terms after pre-processing. If a term is contained in multiple documents across the corpora, it still appears as a single node in the graph. \textit{Metadata nodes} represent a group of tokens, such as a sentence, a tuple, or an attribute. Undirected and unweighted edges connect metadata with the respective data nodes, i.e., a tuple node is connected to its tokens. 

\begin{algorithm}[h] 
\SetAlgoLined
\textbf{Input.} Two sets of documents;\\
G = An un-directed graph\;
    \ForEach{document $doc\_i$ in the first set}
        {G.addNode($doc\_i$) \;
        
        \If{$get\_type(first$ $set)$=table}{
            \ForEach{column $col\_j$ in the document}{
            
            \If{$\neg$ G.hasNode($col\_j$)}{G.addNode($col\_j$)}}
        }
        \If{$get\_type(first$ $set)$=structured}{
        
        parent $\gets$ get parent of document i \;
        \If{G.hasNode(parent)}{G.addEdge($doc\_i$,parent)}
        }
        terms $\gets$ get list of terms in document i \;
        \ForEach{term $tm\_k$ in terms}
        {G.addNode($tm\_k$)\\ 
        G.addEdge($doc\_i$,$tm\_k$) \\
        \If{$get\_type(set)$=table}{
        G.addEdge($col\_j$,$tm\_k$)}}
    }    
    \ForEach{document $doc\_i$ in the second set}
    {G.addNode($doc\_i$)\;
    
    terms $\gets$ get list of terms in the document\;
    \ForEach{term $tm\_j$ in terms}{
    \If{G.hasNode($tm\_j$)}{G.addEdge($doc\_i$,$tm\_j$)}
    
    }
    }
    
\textbf{Output.} Graph G

\caption{Graph Creation}
\label{alg:graphCreation}
\end{algorithm}
 
The graph creation is presented in Algorithm~\ref{alg:graphCreation}. The algorithm takes as input two corpora of pre-processed documents. 
It creates a metadata node for each document in the first corpus (lines 3-4). For example, Figure~\ref{fig:exampleGraph} shows metadata nodes \textit{t1} and \textit{t2} for the tuples. If the document is a table, it also creates a metadata node for every attribute (lines 5-10), such as nodes  \textit{c2}, \textit{c3}, \textit{c4}, and \textit{c5}. 
\rev{These columns are modeled as nodes as they add 2-hop paths across values from the active domain of an attribute.}
 If the document is a structured text, its nodes are modeled as metadata nodes and edges are added to represent relations (lines 12-15).
For each term associated to the document (metadata) node, the algorithm then creates term nodes (lines 18-20) and connects each data node to its respective metadata node (lines 21-24). For example, edges are created to connect \textit{t1} to \textit{Shyamalan}, \textit{Willis}, \textit{B.\_Willis}, \textit{PG}, and \textit{Thriller}. 
Next, metadata nodes for the documents in the second corpus are created (lines 27-28), thus adding \textit{p1} and \textit{p2} in Figure~\ref{fig:exampleGraph}. Term nodes for these documents are connected to their metadata nodes (lines 29-34). For example, metadata node \textit{p1} is connected to data nodes \textit{Willis} and \textit{Comedy}.

Algorithm~\ref{alg:graphCreation} creates different metadata nodes based on the input documents. In this example, it outputs metadata nodes to represent tuple, columns, and text. For the other two tasks (text to text and text to structured text), only text metadata nodes are produced.

\subsection{Connecting metadata nodes} \label{subsec:connections}
The graph creation algorithm never connects metadata nodes from different corpora, as we assume that these connections are hard to infer and are indeed the results of the downstream task in our system. 
\rev{However, those relations can be provided by the users and consumed by our algorithm.}
On the other hand, metadata text nodes from the same structured document can be connected.
For example, for the {taxonomy} in Figure~\ref{fig:kpmg}, the corresponding graph connects metadata text nodes for \textit{Audit programme} and \textit{ISO 19001}. This edge represents the hierarchical relation between the concepts. 

\subsection{Filtering nodes}\label{subsec:summerize}
The graph can become extremely large with real text corpora. This is a problem in terms of performance both for the execution time and for the quality, as it may lead to model several terms that do not contribute to the final matching tasks.
To address this problem, we filter out irrelevant terms in the graph creation.
Algorithm~\ref{alg:graphCreation} does not create data nodes for all terms in both corpora. It starts by creating data nodes for documents in the corpus with the smaller number of distinct tokens and filters out from the second corpus the terms that are not already in the graph.

As our goal is to model the connections across the two corpora, we compromise the loss of some words (and possibly relationships) in the second corpus to focus the learning in the next step on the terms that create bridges between metadata nodes. To limit the loss of possibly relevant words and to create a more compact graph, we present next some techniques to merge token nodes across corpora. 

\begin{figure}[t]
\centering
  \includegraphics[width=0.4\textwidth]{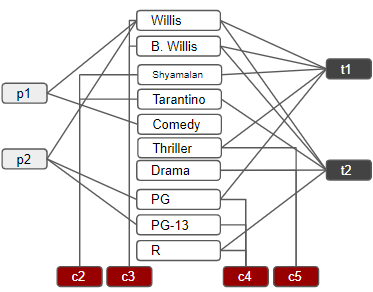}
  \caption{Graph with a sample of the nodes for Example 1.}
  \label{fig:exampleGraph}
  \vspace{-2ex}
\end{figure}

\subsection{Merging nodes}
\label{subsec:external}
Intuitively, (correctly) merging data nodes increases the connectivity between related metadata nodes across corpora and ultimately improves the matching tasks. For example, merging data nodes \textit{Bruce Willis} and \textit{B. Willis} in Figure~\ref{fig:movies} decreases the distance in many paths connecting metadata nodes \textit{P1} and \textit{t1}.
Merging data nodes is easier that solving the metadata matching task and there are several resources available for this operation.
For this matter, we use different techniques to merge data nodes:
\begin{itemize}
    \item \textit{Stemming} merges different forms of a word. For example, in Figure~\ref{fig:kpmg}, stemming merges \textit{planning} from the first paragraph with node \textit{Plan} from \textit{Plan Do Check Act Steps}. 
    \item \textit{Bucketing} 
    merges data nodes with numeric values by using equal width binning and the \textit{Freedman–Diaconis} rule~\cite{freedman1981histogram} to compute buckets' width.
    \item For merging synonyms, acronyms, and typos we use external resources, \rev{such as WordNet}~\cite{miller1998wordnet}. 
\end{itemize}

For the last case, we merge two data nodes if the cosine similarity between their embedding vectors is higher than a threshold $\gamma$. For calculating $\gamma$, we use a list of 17K synonym terms from \textit{WordNet} and define $\gamma$ as the average cosine similarity between their vectors in the pre-trained model that we use for merging. Specifically, for Wikipedia2Vec~\cite{yamada2016joint} we identify and set $\gamma=0.57$.
This approach is widely used in tasks such as {entity linking}~\cite{chen2019yelm} 
and retrieval~\cite{gerritse2020graph}.

\subsection{Tokens and terms} 
One important aspect in graph creation is handling multi-tokens data nodes. There are a lot of meaningful multi-tokens words in document (e.g., movie names) and information is lost if they are split over different single-word data nodes in the graph, e.g., \textit{The Sixth Sense} split over data nodes \textit{Sixth} and \textit{Sense} with \textit{The} filtered out as stop word. A possible solution for tables is to represent the whole cell value as a single data node (\textit{The\_Sixth\_Sense}). But this granularity has also drawbacks as it may lead to graphs that miss important connections across corpora. For example,  
if \textit{B. Willis} in the review and node \textit{Willis} are not merged, a strong connection between the correct metadata nodes is lost.

We use a combination of the two approaches that solves both problems. For each text in a corpus, we generate possible $n$-gram tokens for $n=1,\ldots,n$. For example, for $n=3$, the graph represents \textit{The Sixth Sense} using five data nodes: \textit{Six}, \textit{Sense}, \textit{The\_Six}, \textit{Six\_Sense}, and \textit{The\_Six\_Sense}. This increases the chance of connecting terms of the second corpus with nodes from the first corpus. We identified the default value of $n$ for every scenario by profiling a file of titles of \textit{Wikipedia} articles. About $99\%$ of the titles have at most three tokens. 
Experiments in Section~\ref{subsec:impacts} show that increasing $n$ up to three improves the quality performance but  
we observe diminishing return with higher values.
Other results also show that by increasing $n$ in character tokenization, the lexicon size grows rapidly and precision diminishes for most languages~\cite{mcnamee2004character}.

\section{Graph Expansion and Compression} 
\label{sec:expand}
By generating the graph with Algorithm~\ref{alg:graphCreation}, the paths between metadata nodes represent the relationships which are present in the documents and tables. This network of connections leads to embeddings that ultimately guide the metadata matching process. However, the data relationships are not all the existing relations between two real objects represented in the graph. Real entities and concepts are connected by more relationships that are missing from the corpus at hand. 

For example, an actor and a director may have worked together in a movie that is not in the movie table or in any of the reviews. 
Such external relationships can be very valuable if represented in our graph as they lead to better embeddings and ultimately enable better matching. However, while expanding the graph is a valid solution to include external information, we should be careful in trying to remove useless or even misleading new nodes and edges in order to keep the graph as little as possible in terms of size. We present solutions to address these two tasks in this section.

\subsection{Expanding the graph with external information}

A natural approach to expand the generated graph is to employ external resources \rev{that model information with nodes and edges}, such as ontologies and knowledge bases. By exploiting existing resources, we add information
to the graph. 

For example, in Figure~\ref{fig:exampleGraph}, \textit{p1} is the review related to tuple \textit{t2}. Even though there are seven paths between these two metadata, only one of them has three or less nodes: {p1 $\rightarrow$ Willis $\rightarrow$ t2}.
By expanding this graph with new nodes and edges, we can add new meaningful paths between these nodes, improve their embeddings, and increase their chance of being matched. 
Consider as a resource of external information the knowledge graph DBpedia~\cite{bizer2009dbpedia}. Among the relations for entity \textit{Tarantino} in DBpedia, there is the following triple: \textit{style(Tarantino, Comedy)}. Adding this new edge to the graph creates nine new path between \textit{p1} and \textit{t2} including one with less than three nodes: {p1 $\rightarrow$ Comedy $\rightarrow$ Tarantino $\rightarrow$ t2. }

\begin{figure}[t]
\centering
  \includegraphics[width=0.45\textwidth]{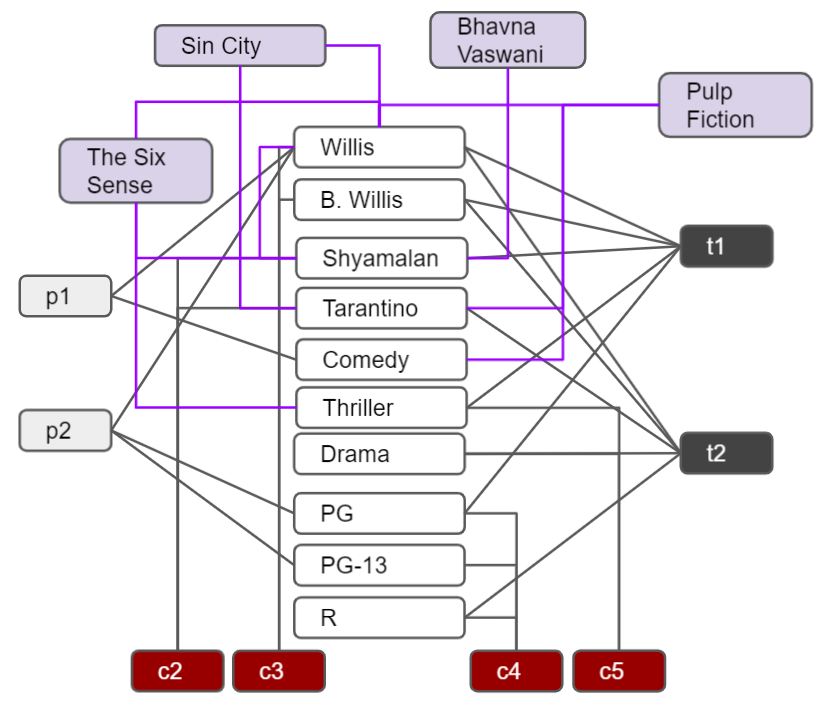}
  \caption{Expanded graph for Example 1.}
  \label{fig:expandedGraph}
  \vspace{-2ex}
\end{figure}

Different external resources can be exploited to expand a graph. In graphs with named entities, we are interested in finding more information about them, such as data about their spouse, country, university, workplace, etc. This information can be extracted from existing entity-centric knowledge bases (KBs) such as \textit{DBpedia} and \textit{Wikidata}. For example, the graph presented in Figure~\ref{fig:exampleGraph} contains information about movies and their casts, those are entities for which we can use a KB for expansion. By expanding this graph with DBpedia, we enrich it with new edges such as \textit{starringOf(Willis, Pulp Fiction)} and \textit{spouse(Shyamalan, Bhavna Vaswani)} as shown in Figure~\ref{fig:expandedGraph}. 

Textual corpora do not contain only named entities, but also concepts, generic nouns and verbs. For example, by expanding the word \textit{management} in Figure~\ref{fig:kpmg}, we connect it with the relevant words in the correct text paragraph, such as \textit{planning}. 
In these cases, other external resources can be exploited, such as 
\textit{ConceptNet}~\cite{SpeerCH17} and \textit{Wordnet}.

\begin{algorithm}[h] 
\SetAlgoLined
\KwIn{Un-directed graph G }
\myinput{External resource E }

\tcp{\footnotesize\ttfamily{Expanding by fetching connections from E }}
    
    \ForEach{node in G}{
        \If{node is not a metadata node}{
        relations $\gets$ all connections of node in $E$\;

            \ForEach{(node,m) in relations}{
                \If{$\neg$ G.hasNode(m)}{
                G.addNode(m) \\}
                G.addEdge(node,m)                            }
        }

    }
\tcp{\footnotesize\ttfamily{Cleaning the graph }}
    \ForEach{node in G}{
        \If{degree(node) == 1}{G.removeNode(node)}}

\textbf{Output.} Expanded graph G

\caption{Graph Expansion Algorithm}
\label{alg:graphExpansion}
\end{algorithm}

Algorithm~\ref{alg:graphExpansion} shows how we exploit any external resource to fetch all connections for every data node in the graph. We also remove any sink node, i.e., nodes that are not connected to more than one other node. E.g., in Figure~\ref{fig:expandedGraph} node \textit{Bhavna Vaswani} is only connected to \textit{Shyamalan} and can be removed. 

The expansion technique introduces new graph paths that affect the matching process between metadata nodes. For example, in the original graph in Figure~\ref{fig:exampleGraph}, there is one path with less than five nodes between metadata nodes \textit{p1} and \textit{t2}, but the their shortest path is of size two after expansion. 
    
\subsection{Pruning nodes and edges for compression}
Expansion introduces relevant connections between related nodes, but it also increases its size by adding  new nodes and edges that are not helpful for our tasks.
For example, there are more than 800 relations for entity \textit{Quentin Tarantino} in DBpedia but only a few of them increase the chance of matching \textit{p1} and \textit{t2} (e.g., \textit{directorOf(Quentin Tarantino, Pulp Fiction)}, \textit{redirectsOf(Quentin Tarantino, Samuel Jackson)}).
As new nodes and relationships increase the execution time for random walks and embedding generation, we should avoid keeping nodes and edges that do not contribute to the connections among metadata nodes. 

We therefore introduce a graph compression techniques to reduce the size of our graph after the expansion phase. Compression for static graphs has been studied for its benefits in terms of reduction of data volume and storage, which in turn enables speedup of algorithms and queries~\cite{adamic2001search}. Noise elimination has also been reported as an important effect of the compression, with the removal of erroneous nodes and labels~\cite{LiuSDK18}. Compression methods can be based on node sampling~\cite{adamic2001search, stumpf2005subnets, leskovec2006sampling}, edge sampling~\cite{krishnamurthy2005reducing, ahmed2013network}, or exploration based sampling~\cite{leskovec2005graphs, rezvanian2015sampling, li2019walking}. 
Most methods are configurable w.r.t. the desired compression ratio, i.e., the desired size of the output graph compared to the input graph. 

As these methods are very general, they are not application specific and cannot make use of the node types in our graph.
Our experiments in Section~\ref{sec:experiments} show that these techniques can help in reducing the size of the graph by filtering nodes, but do not preserve  
good performance in the matching task. 

Our key observation is that the goal of our graph and our embeddings it to match metadata nodes. A crucial component in determining the  distance between the embeddings for two metadata nodes in their distance in the graph. We therefore start the design of our compression algorithm from the idea that it should preserve the shortest path across all the metadata nodes in the two corpora. This is a quadratic number of paths w.r.t. the number of metadata nodes, but this is unavoidable as we do not know at compression time what are the metadata to match.

Inspired by an existing graph compression technique that exploits shortest paths, namely \textit{SSP}~\cite{rezvanian2015sampling}, we introduce an algorithm tailored at our graph and matching application.
The original \textit{SSP} is an exploration based sampling method which takes a sampling size as input. It randomly picks a pair of nodes in each iteration, computes their shortest path, and adds nodes and edges of the shortest path to the output graph.
In our setting, the idea is to use metadata nodes in distinct corpora for the selection of node pairs. This guarantees that metadata nodes are connected and keeps in the graph the data nodes that are modeling their relationship concisely.

\begin{algorithm}[h] 
\SetAlgoLined
\KwIn{Un-directed graph G }
\myinput{Compression ratio $\beta$ }
CG = empty un-directed graph \\

i = 0 \\
L = $\beta$ * size(G.nodes())

\While{i $<$ L}{
\tcp{\footnotesize\ttfamily{Select two random metadata nodes}}

first $\gets$ a random node from the first corpus \\
second $\gets$ a random node from the second corpus \\

shortest\_paths $\gets$ find all shortest paths between \textbf{first} and \textbf{second} in G
    
\tcp{\footnotesize\ttfamily{Add nodes and edges of the paths to CG}}

    \ForEach{path in shortest\_paths}{

    CG.add(path)

    }
    i += 1\\

}

\textbf{Output.} Compressed graph CG

\caption{Graph Compression MSP}
\label{alg:ssp}
\end{algorithm}

Algorithm~\ref{alg:ssp} shows our graph compression based on the idea of using Metadata Shortest Path (\textit{MSP}). It takes an un-directed graph and a compression ratio $\beta$ as input and returns a compressed graph. We define the number of iterations of the Algorithm by multiplying $\beta$ and the number of nodes in the graph. We also make sure that all metadata nodes, even if not sampled at lines 6-7, are connected to the graph with at least one shortest path.

There is however an orthogonal challenge in such an aggressive compression. Keeping only shortest paths among metadata nodes may lead to a graph with similar embeddings for all nodes. Assume that all shortest paths are of length three and all of them pivot on a single, very popular data node. This extreme situation leads to all metadata nodes ending up with the same embedding, therefore making the matching process impossible. This observation highlights that compression is not suitable in all cases. We discuss in the experiments how we can recognize setting that are less likely to benefit from the compression.

\section{Matching Text and Structured Data} \label{sec:secondStep}

In this section we  
 discuss how to generate and compare node embeddings for the unsupervised matching of objects.

\subsection{Embeddings Generation} \label{subsec:randomwalk}
We generate 
embeddings for the graph nodes. Multiple methods can be employed to generate embeddings directly from the graph~\cite{hamilton2017representation,perozzi2014deepwalk}. A less resource intensive solution is to use word embedding models on walks over the graph~\cite{grover2016node2vec}. In our default setting, we use the second approach  
as we found the results with different methods comparable in quality, but the latter is faster and less demanding in terms of resources. 

\begin{algorithm}[h] 
\SetAlgoLined
\KwIn{Graph G }
\myinput{Number of RWs $n$ }
\myinput{Length of RWs $l$ }

docs = [] \\

\For{i in range(0,$n$)}{
\For{node in G}{
rw = [] \\
\While{len(rw) $< l$}{
next $\gets$ a random neighbor of node \\
rw.append(next.label())
}
docs.append(` '.join(rw))
}
}
M $\gets$ Word embedding model \\
M.train(docs)

\textbf{Output.} Embeddings for data and metadata nodes

\caption{\rev{Embedding Generation}}
\label{alg:embgen}
\end{algorithm}

\rev{Algorithm~\ref{alg:embgen} shows the process of generating embeddings. The output of this algorithm are word embeddings that map words in the random walks (data and metadata nodes) to vectors of real numbers. In a word embedding model, words with more co-occurrences in sentences show smaller distances in the corresponding vectors. In our model, related metadata nodes have a higher chance of appearing together in a random walk, thus their vectors show smaller distances.} 
In our default method, a random walk starts from every graph node and at each step it randomly chooses the next node among the current node's neighbors. A sentence is derived with the concatenation of nodes traversed by the walk. The union of the sentences is then processed with \textit{Word2Vec} or similar methods.
Multiple parameters of the random walks affect the quality of the embeddings, including whether random walk should be limited to some nodes or not, the length of a random walk, and how many random walks should be generated for every graph node.
In Section~\ref{subsec:impacts}, we report the impact of these parameters on the quality of the proposed model. 

We stress the importance of the unified graph and of the walks in our context.
While any data structure can be serialized as a sequence of sentences (e.g., row by row), this is not effective in practice for learning embeddings for two reasons. First, the resulting sentences do not follow meaningful patterns as in real languages; the relationships that are nicely captured in real text are missing. This is especially true for transformer based solutions based on attention. Second, a simple serialization misses the structural dependencies in a relation; existing features in the data are not exploited. Given text documents and a relation, our graph and walks enable the joint representation for the given corpora.

\subsection{Matching Metadata Nodes} 
\label{subsec:match}
In the final step, 
the input of the matching module is a metadata node for a document in the first corpus and the metadata nodes representing all documents in the second corpus. The embedding vectors are used to match such nodes and ultimately the documents they represent (text paragraphs, text sentences, or tuples). The distance between two nodes' vectors is used for matching them. 
Given the embeddings, we use cosine similarity to identify the top-$k$ neighbours in the second corpus for the metadata node from the first corpus. 

Differently from the graph creation, in the metadata matching we found more effective to  start the process from the larger corpus and this is our default configuration. However, this decision can be changed according to the specific task. For example, in text matching we do it claim by claim (from the smaller corpus) as this is the natural setting for the application.

\section{Experiments} \label{sec:experiments}
We first introduce our execution setting and the baseline methods. We then report the results in three matching tasks: i) text to data, ii) text to structured text, iii) text to text. We do not
report performance for the data to data task since it has already been studied in previous work~\cite{CappuzzoPT20}. 
Finally, we report execution times and discuss the impact of the different parameters and optimizations in our solution. Code and datasets
are available at \url{https://github.com/naserahmadi/TDmatch}.

\vspace{1ex}
\noindent \textbf{Execution setting.} Experiments have been conducted on a laptop with CPU Intel i5-7300U, 4x2.6GHz cores and 8GB RAM. 
For fine tuning the 
baselines, we used a \textit{Google Colaboratory} instance with CPU Intel 2x2.20GHz, 13GB RAM, and NVIDIA Tesla T4 GPU (16GB memory). Algorithms are written in \textit{Python} with the \textit{Numba} compiler.

\vspace{1ex}
\noindent \textbf{Baselines.} 
We compare our approach to several baselines; 
we report only the best performing baselines for every task. 
Some baselines rely on \textit{training} over the given documents and data (as our approach), while others use \textit{pre-trained} resources. We further distinguish unsupervised and supervised solutions. 

For unsupervised methods based on \textit{training}, we test \textit{Word2Vec} (\wvec) for word embedding and \textit{Doc2Vec} (\dvec) for document embedding. We obtain embeddings from the documents at hand and then use such embeddings to identify matches.
We generate embeddings for longer texts with the mean of the vectors of their tokens~\cite{de2016representation}. We use vectors of size 300, Skip-Gram for \textit{Word2Vec} and 
DBOW for \textit{Doc2Vec}.
For unsupervised approaches using \textit{pre-trained} embeddings,  
we report on {\textit{SentenceBERT}} (\sBert).
Unsupervised methods match objects with the algorithm in Section~\ref{subsec:match}.

We denote supervised methods using pre-trained models with $^*$ for clarity and always report results for 5-fold cross validation. 
The first approach is based on fine-tuning for a multi-label 
classification task on 
\textit{BERT} large (\lBertS). 
We also report the results for two supervised state-of-the-art entity matching methods for the text to data task: \textit{Ditto}~\cite{DITTO21} and \textit{DeepMatcher}~\cite{mudgal2018deep}. These methods take two tables as input and compute the matching probability for tuples from different tables. We represent text documents as tuples of a table with one attribute. 
\rev{We report the results for text-to-data matching also using \textit{TAPAS}~\cite{HerzigNMPE20}, a BERT-based  model designed and pre-trained for answering questions about tabular data.}
We use 60\% of the annotated data to train these models. 
Finally, we report for 
\textit{Reranking} (\rankS), a supervised algorithm that learns to rank using a pairwise loss~\cite{ShaarBMN20}.

For our unsupervised approach, we used \textit{Word2vec} (\rwWord) on the random walks (RW) generated on the graph. 
In the default configuration, we generate 100 random walks of length 30 for every node. For the text to data task, we use \textit{Skip-gram} with a window of size three as in the data to data match~\cite{CappuzzoPT20}, while for text oriented tasks we use \textit{CBOW} with a window of size 15. {We report for our method with (\rwWordEx) and without (\rwWord) applying the expansion technique in Section~\ref{sec:expand}. We use ConceptNet as our default external resource for expanding graphs, except for \textit{IMDB} where we employed DBpedia as this relation contains mostly entities.}

\subsection{Text to Data}
For the \textit{text to data} matching we use two datasets.
We created a first scenario from the Internet Movie Database (\textit{IMDb}) website with a corpus of movies reviews and a database of movies. We also report results for the \textit{CoronaCheck} scenario, which matches COVID-19 claims to the official datasets~\cite{KaragiannisSPT20}. For both scenarios, the task is to find tuples related to each sentence. For example, a sentence ``Number of cases in US is higher than China'' required to match two rows of a table to verify the claim.

\newcounter{num}
\def\Num{\par\smallskip\refstepcounter{num}{\bfseries\arabic{num}}.\hspace{0.2em}}
\def\num{\par\medskip\refstepcounter{num}\hangindent2em{\bfseries\arabic{num}}.\hspace{1em}}
\newenvironment{Numera}
{\parindent0pt\par\smallskip}
{\setcounter{num}{0}\par\medskip}

\vspace{1ex}
\noindent \textbf{Datasets.}  
As the task is novel, we release two new scenarios:
\begin{Numera}
    \Num {\bf IMDb.} We created the dataset by manually matching two reviews for every movie in ``top 1K of all times" to a sample of 50k tuples from the official IMDb dataset.
    The 2k reviews contain one to 207 sentences, sixteen on average.
 We created two versions of the target relation: an easier one with 13 attributes, including the title information (\textbf{WT}) and a more challenging one without title (\textbf{NT}). 

    \Num {\bf CoronaCheck.} This scenario contains a corpus of sentences about COVID-19 spread and effects, such as daily total death cases and new confirmed monthly cases, annotated w.r.t. the corresponding tuples in a dataset with 1.2k tuples about daily cases for all countries. 
    We report for a dataset with 7k sentences created from the data (\textbf{Gen}) and a more challenging dataset with 50 sentences submitted by users on the website \url{https://coronacheck.eurecom.fr} (\textbf{Usr})~\cite{KaragiannisSPT20}. 
\end{Numera}

\begin{table}
\footnotesize
\caption{\rev{Quality of match results for \textit{IMDb} scenario.}}
\label{tab:imdb}
\begin{tabular}{ccc|ccc|ccc}

& Method   & MRR  & \multicolumn{3}{c|}{MAP@k} & \multicolumn{3}{c}{HasPositive@k}  \\ 
 
&   & & 1 & 5 & 20 & 1 & 5  & 20 \\ 
\hline
\multirow{6}{*}{\rotatebox{90}{WT}} 
& \sBert & .254 & .088 & .142 & .159 & .171 & .339 & .510 \\
&  \rwWord  & .853 & .400 & .678 & .682 & .802 & .919 & .942 \\ 
&  \rwWordEx  & .\textbf{868} & .\textbf{410} & .\textbf{691} & .\textbf{706} & .\textbf{820} & .\textbf{926} & .\textbf{955} \\ 

\cline{2-9}
& \rankS  & .535 & .218 & .351 & .376 & .438 & .645 & .797 \\
& \ditto  & .759 & .349 & .549 & .553 & .699 & .839 & .877 \\
& \rev{\tapas} & .722 & .375 & .525 & .526 & .802 & .849 & .929 \\

\hline
\hline

\multirow{6}{*}{\rotatebox{90}{NT}}   

& \sBert & .218 & .067 & .118 & .139 & .136 & .301 & .454 \\
&  \rwWord  & .780 & .362 & .574 & .589 & .727 & .841 & .906 \\
&  \rwWordEx  & .\textbf{792} & .\textbf{371} & .\textbf{587} & .\textbf{598} & .\textbf{749} & .\textbf{854} & .\textbf{911}    \\   

\cline{2-9}
& \rankS  & .404 & .156 & .236 & .260 & .312 & .494 & .688 \\
& \ditto  & .560 & .265 & .386 & .410 & .428 & .689 & .814 \\
& \rev{\tapas} & .643 & .327 & .354 & .358 & .678 & .706 & .715 \\

\end{tabular}
\end{table}

\begin{table}
\centering
\footnotesize
\caption{\rev{Quality of match results for \textit{CoronaCheck} scenario.}}
\label{tab:corona}
\begin{tabular}{ccc|ccc|ccc}
& Method   & MRR  & \multicolumn{3}{c|}{MAP@k} & \multicolumn{3}{c}{HasPositive@k}  \\ 
&    &      & 1    & 5    & 20 & 1    & 5    & 20 \\ 
\hline
\multirow{7}{*}{\rotatebox{90}{Gen}} 
& \sBert  & .486 & .294 & .463 & .483  & .295 & .752 & .916 \\
&  \rwWord  & .728 & .575 & .718 & .725  & .578 & .945 & .995 \\
&  \rwWordEx  & .\textbf{755} & .\textbf{601} & .\textbf{746} & .\textbf{752}  & .\textbf{611} & .\textbf{959} & .\textbf{996} \\ 
\cline{2-9}
& \rankS  & .460 & .287 & .438 & .455 & .289 & .703 & .845 \\
& \deep  & .376 & .347 & .368 & .374 & .349 & .395 & .439 \\
& \ditto  & .160 & .030 & .161 & .203 & .066 & .283 & .518 \\
& \rev{\tapas} & .394 & .325 & .367 & .389 & .330 & .452 & .723 \\

\hline
\hline
\multirow{7}{*}{\rotatebox{90}{Usr}} 
& \sBert  & .354 & .177 & .284 & .320 & .200 & .620 & .860  \\
& \rwWord   & .518 & .296 & .427 & .472 & .306 & .755 & .979  \\
& \rwWordEx   & .\textbf{538} & .\textbf{329} & .\textbf{451} & .\textbf{496} & .\textbf{371} & .\textbf{771} & \textbf{1} \\

\cline{2-9}
& \rankS  & .332 & .137 & .256 & .303 & .160 & .600 & .880 \\
& \deep  & .321 & .200 & .200 & .248 & .280 & .280 & .600 \\
& \ditto  & .153 & .020 & .100 & .123 & .040 & .281 & .407 \\
& \rev{\tapas} & .192 & .053 & .053 & .077 & .053 & .053 & .474 \\

\end{tabular}
\end{table}

\vspace{1ex}
\noindent \textbf{Evaluation Measures.}  
Mean
Reciprocal Rank (\textit{MRR}) is the average of reciprocal ranks of queries, i.e., the multiplicative inverse of the rank of the first correct answer. 
Mean Average Precision
({MAP}) is the mean of the precision scores after each relevant document is retrieved and we report MAP truncated at rank $k$ (\textit{MAP@k}).
We also report HasPositive@k for determining whether there is a true positive among the top-$k$ results.

\vspace{1ex}
\noindent \textbf{Matching results.}
As the training-based methods (\wvec, \dvec) do not take tables as input, we 
serialize every tuple to a sentence using two special tokens (\textit{[COL]} and \textit{[VAL})~\cite{DITTO21}, e.g., the first row in Figure~\ref{fig:movies} starts with ``[COL] title [VAL] The Sixth Sense [COL] director [VAL] Shyamalan". We then generate an embedding vector for every resulting sentence and match vectors for tuple and text metadata nodes. 
As results are poor for these baselines, we do not report them.
For the pre-trained models, we report for \sBert, 
\rankS, \ditto, \deep, and \rev{\tapas}. 

Table~\ref{tab:imdb} and Table~\ref{tab:corona} show the results on the \textbf{IMDb} and \textbf{CoronaCheck} scenarios, respectively. 
Our method outperforms unsupervised \sBert in all scenarios and the techniques in Section~\ref{sec:expand} show a positive effect in both datasets. 
For \textbf{IMDb}, we observe an absolute increase of 0.45 for MRR with \rwWord in both datasets and at least 410x relative improvement for Positive@1. In \textbf{CoronaCheck}, the increase for \textbf{Gen} sentences is an absolute 0.24 for MRR and up to 0.30 for MAP@k and Positive@k. For the \textbf{Usr} sentences, increases are up to 0.2 for MRR and MAP and up to 0.18 for Positive@k.     
Our model clearly outperforms also supervised methods. Results show that pre-trained models fail short in this task and that the joint modeling enabled by our graph is needed to achieve good matches. 
We do not report \textit{DeepMatcher} on \textit{IMDB} because it failed due to the limited amount of memory in our machine.

\subsection{Text to Structured Text} 
\label{sec:auditExp}
In this task, we match taxonomy elements to a text document in a real enterprise scenario from an auditing company.

\vspace{1ex}
\noindent \textbf{Dataset.} This scenario contains 1622 audit text documents (containing one to 17 sentences, three on average) and a taxonomy containing 747 auditing concepts.  
Each path spans multiple nodes, e.g., 
$r_1: a \rightarrow b \rightarrow c \rightarrow d$, where each variable is a concept.
Right arrows show the hierarchical relations between terms, e.g., in $r_1$ c is a child of b and b is a child of a. The length of taxonomy paths are between two and five nodes (four on average). The final graph has 5.9k nodes and 164k edges. Text documents are manually 
matched to concept nodes by domain experts. 
About $40\%$ of documents 
are annotated with one concept, $10\%$ are matched to two concepts, and the rest are matched with three to $27$ concepts (four on average).

\begin{table}[h]

	\centering
		\caption{Exact and Node scores for 
		structured text matches.}
		\small
\vspace{1ex}
\label{tab:multilabel-res}

\begin{tabular}{|c c|c c c|c c c|}
& & \multicolumn{3}{c|}{\textbf{Exact Scores}} & \multicolumn{3}{c|}{\textbf{Node Scores}} \tabularnewline

& \multicolumn{1}{ c| } {\it Method} & {\it P} & {\it R} & {\it F} & {\it P} & {\it R} & {\it F} \\
\hline

\parbox[t]{2mm}{\multirow{4}{*}{\rotatebox[origin=c]{90}{K=1}}} & \multicolumn{1}{ |c| }{\dvec} & .254 & .217  & .234  & .554 & .503 &  .527 \\ \cline{2-8}

& \multicolumn{1}{ |c| }{\sBert} & .094 & .071 & .081 & .379 & .358 & .368 \\ \cline{2-8}

& \multicolumn{1}{ |c| } {\rwWord} & .346 &  .265 & .300  & .593  &  .530 & .560   \\ \cline{2-8}

& \multicolumn{1}{ |c| }{\rwWordEx} & .\textbf{367} &  .\textbf{282} & .\textbf{319} & .\textbf{601} & .\textbf{545} & .\textbf{572}  \\ \cline{2-8}

\hline 
\hline

\parbox[t]{2mm}{\multirow{2}{*}{\rotatebox[origin=c]{90}{K=1}}} & \multicolumn{1}{ |c| }{\rankS} & .162 & .125 & .138 & .425 & .392 & .408 \\ \cline{2-8}

& \multicolumn{1}{ |c| }{\lBertS} & .{381} &
.{304} & {.{338}} & .{626} & .{567} & {.{595}} \\ \cline{2-8}

\hline
\hline
\hline

\parbox[t]{2mm}{\multirow{4}{*}{\rotatebox[origin=c]{90}{K=3}}} & \multicolumn{1}{ |c| }{\dvec} & .176 & .386  & .242  & .485 & .564 & .521  \\ \cline{2-8}

& \multicolumn{1}{ |c| }{\sBert} & .065 & .014 & .088 & .362 & .431 & .393 \\ \cline{2-8}

& \multicolumn{1}{ |c| } {\rwWord} & 201 & .434  & .275  & .521  & .652  & .579   \\ \cline{2-8}

& \multicolumn{1}{ |c| }{\rwWordEx} & .\textbf{214} & .\textbf{475}  & .\textbf{295} & .\textbf{528} & \textbf{.670}  & .\textbf{594}  \\ \cline{2-8}

\hline 
\hline

\parbox[t]{2mm}{\multirow{2}{*}{\rotatebox[origin=c]{90}{K=3}}}& \multicolumn{1}{ |c| }{\rankS} & .162 & .125 & .138 & .425 & .392 & .408 \\ \cline{2-8}

& \multicolumn{1}{ |c| }{\lBertS} & .183 & .417 & .254 & .487 & .{678} & .566 \\ \cline{2-8}

\hline 
\hline
\hline

\parbox[t]{2mm}{\multirow{4}{*}{\rotatebox[origin=c]{90}{K=5}}} & \multicolumn{1}{ |c| }{\dvec} & .132 & .470  & .206  & .457 & .679 & .546  \\ \cline{2-8}

& \multicolumn{1}{ |c| }{\sBert} & .052 & .179 & .080 & .356 & .473 & .406 \\ \cline{2-8}

& \multicolumn{1}{ |c| } {\rwWord} & .145 & .508  & .222  & .478 &  .699 & .568   \\ \cline{2-8}

& \multicolumn{1}{ |c| }{\rwWordEx} & .\textbf{151} & .\textbf{533} & .\textbf{236} & .\textbf{485} & .719 & .\textbf{580}  \\ \cline{2-8}

\hline 
\hline

\parbox[t]{2mm}{\multirow{2}{*}{\rotatebox[origin=c]{90}{K=5}}} & \multicolumn{1}{ |c| }{\rankS} & .072 & .242 & .110 & .365 & .522 & .429 \\ \cline{2-8}

& \multicolumn{1}{ |c| }{\lBertS} & .135 & .508 & .213 & .446 & .{740} & .556 \\ \cline{2-8}

\hline 
\hline
\hline

\parbox[t]{2mm}{\multirow{4}{*}{\rotatebox[origin=c]{90}{K=10}}} & \multicolumn{1}{ |c| }{\dvec} & .087 &  .587 & .152  & .42 & .758 &  .541 \\ \cline{2-8}

& \multicolumn{1}{ |c| }{\sBert} & .038 & .253 & .066 & .347 & .541 & .423 \\ \cline{2-8}

& \multicolumn{1}{ |c| } {\rwWord} & .092 &  .613 & .160  & .437  & .768  &  .557  \\ \cline{2-8}

& \multicolumn{1}{ |c| }{\rwWordEx} & .\textbf{094} & .\textbf{629}  & .\textbf{164} & .\textbf{438} & \textbf{.783} & .\textbf{562}  \\\cline{2-8}

\hline 
\hline

\parbox[t]{2mm}{\multirow{2}{*}{\rotatebox[origin=c]{90}{K=10}}} & \multicolumn{1}{ |c| }{\rankS} & .051 & .324 & .088 & .350 & .592 & .440 \\ \cline{2-8}

& \multicolumn{1}{ |c| }{\lBertS} & .081 & .584 & .141 & .393 & .{797} & .526 \\ \cline{2-8}

\hline

\end{tabular}
\end{table}

 \vspace{1ex}
\noindent \textbf{Evaluation Measures.} 
For this task, we change quality measures as we show results at different granularity.
We report Precision, Recall and F-score for concepts (in the taxonomy) assigned to every document w.r.t. the ground truth. As different taxonomy nodes can contain the same text, we compare the root to node path in the measures.  
With \textbf{Exact} matches, we consider a match in the top-k valid only if it is \textit{equal} to the path in the ground truth.
As two paths can be partially overlapping, we consider also partial matches with the \textbf{Node} score, which measures the intersection between the matched path(s) and the closest path(s) in the ground truth. For an accurate calculation, we exclude two most general levels of the taxonomy (root and first level under it) in the intersection and denote the new path with $p'$. We then use formula (\ref{eqn:node_level}) below to calculate the \textbf{Node} score for two paths \textit{$p_1$} and $p_2$. 

\vspace{-2ex}
\begin{equation}
\small
 \label{eqn:node_level}
 Node(p_1, p_2) = 
 \frac
 { intersection((nodes(p'_1), nodes(p'_2) )}
 {maximum ((nodes(p'_1), nodes(p'_2))  } 
 \end{equation}

\vspace{-0.5ex}
 Consider $r_1$ and $r_2: a \rightarrow b \rightarrow c$. After excluding the general nodes, we obtain $r_1: c$ and $r_2: c \rightarrow d$, thus 
 Node($r_1$,$r_2$) = 0.5.

\vspace{1ex}
\noindent \textbf{Matching results.}
Table~\ref{tab:multilabel-res} reports for both measures the precision, recall and F-score for matching top-$k$ paths to every document for different $k$ values. 
Results show that the task is very difficult. Indeed, different auditors have different opinions about the right matches for a given taxonomy node and the ground truth is constructed after a discussion to reach consensus. In this hard task, our methods outperform unsupervised methods with a large margin. This scenario contains some domain specific terms that are not covered by pre-trained models as we can observe by \dvec (trained on the audit data) outperforming unsupervised \sBert. Only for the top-1 case the supervised \textit{BERT\_large} shows small margins for both measures. Supervised classifiers are effective for documents matched against one concept but do not have enough training data for the other cases.
In Section~\ref{subsec:improve}, we show how our model combined with \sBert outperform 
supervised \textit{BERT\_large} 
solution also for k=1.

\subsection{Text to Text}
We evaluate our framework in matching documents between \rev{three} text corpora. While our solution is tailored towards structured data and text, its results are better than unsupervised state of the art baselines for this task and close to supervised ones. Two datasets come from the task of detecting previously fact-checked claims~\cite{ShaarBMN20}.
Given a check-worthy input claim and a set of verified claims, the goal is to
rank the verified claims that help check the input claim
it, 
above other claims.
\rev{We also test the dataset from the STS (semantic textual similarity) GLUE task~\cite{cer2017semeval} as an unsupervised matching task. In this dataset, an original similarity score between text pairs is defined between 0 (completely dissimilar) and 5 (completely equivalent). We consider two snippets a true match when they have in the ground truth a score equal or greater than \textit{k}.}

\vspace{1ex}
\noindent \textbf{Datasets.}  
The \textit{Snopes} dataset contains a set of 1k claims (tweets) and 11k verified claims (facts), while the \textit{Politifact} dataset contains 768 claims (made by politicians) and 16.6k verified claims (facts). {Text documents contain from one to nine sentences in \textit{Snopes} and from one to 11 in \textit{Politifact}. On average they have less than two sentences.}
We match top$-k$ verified claims (facts) for every claim.
\rev{The \textit{STS} dataset contains a set of 7k pairs of sentences and we report results for thresholds k=2 (5k pairs) and k=3 (3.7k pairs).}

 \vspace{1ex}
\noindent \textbf{Evaluation Measures.}  
We use Mean Reciprocal Rank (\textit{MRR}), Mean Average Precision at k
(\textit{MAP@k}), and HasPositive@k.

\vspace{1ex}
\noindent \textbf{Baselines.} 
We use methods with good results reported for these datasets~\cite{ShaarBMN20}: unsupervised \sBert and supervised \rank.

\begin{table}[t]
\caption{Quality of match results for \textit{Politifact} scenario.}
\centering
\footnotesize
\label{tab:politi}
\begin{tabular}{cc|ccc|ccc}
Method & MRR & \multicolumn{3}{c|}{MAP@k} & \multicolumn{3}{c}{HasPositive@k} \\ 
 &  & 1 & 5 & \multicolumn{1}{c|}{20} & 1 & 5 & 20 \\ \hline
\sBert & .395 & .354 & .372 & \multicolumn{1}{c|}{.382} & .362 & .417 & .496 \\
  \rwWord & .489  & .346 & .396 & \multicolumn{1}{c|}{.401} & .409 &  .579 & .702\\

 \rwWordEx & .\textbf{507}  & .\textbf{358} & .\textbf{406} & \multicolumn{1}{c|}{.\textbf{418}} & .\textbf{429} &  .\textbf{600} & .\textbf{726} \\

 \hline
 \hline
\rankS & .608 & .531 & .588 & \multicolumn{1}{c|}{.599} & .535 & .688 & .787
\end{tabular}
\end{table}

\begin{table}[t]
\caption{Quality of match results for \textit{Snopes} scenario.}
\centering
\footnotesize
\label{tab:snopes}
\begin{tabular}{cc|ccc|ccc}
Method & MRR & \multicolumn{3}{c|}{MAP@k} & \multicolumn{3}{c}{HasPositive@k} \\ 
 &  & 1 & 5 & \multicolumn{1}{c|}{20} & 1 & 5 & 20 \\ \hline
\sBert & .543 & .457 & .527 & \multicolumn{1}{c|}{.535} & .457 & .648 & .724 \\
 \rwWord & .695 & .586 & .688 & \multicolumn{1}{c|}{.693} & .587 & .820 & .886 \\
\rwWordEx & .\textbf{708} & .\textbf{613} & .\textbf{698} & 
\multicolumn{1}{c|}{.\textbf{706}} & .\textbf{614} & .\textbf{843} & .\textbf{898} \\

\hline
\hline
\rankS  & .788 & .691 & .782 & \multicolumn{1}{c|}{.784} & .693 & .894 & .925 
\end{tabular}
\end{table} 

\begin{table}[t]
\caption{\rev{Quality of match results for \textit{STS} scenario.}}
\centering
\footnotesize
\label{tab:sts}
\begin{tabular}{ccc|ccc|ccc}

& Method   & MRR  & \multicolumn{3}{c|}{MAP@k} & \multicolumn{3}{c}{HasPositive@k}  \\ 
 
&   & & 1 & 5 & 20 & 1 & 5  & 20 \\ 
\hline
\multirow{4}{*}{\rotatebox{90}{\rev{k=2}}} 
& \sBert & .739 & .649 & .723 & .733 & .657 & .836 & .920 \\
&  \rwWord  & .780 & .691 & .765 & .772 & .703 & .872 & .947 \\ 
&  \rwWordEx  & .\textbf{796} & .\textbf{707} & .\textbf{785} & .\textbf{788} & .\textbf{716} & .\textbf{887} & .\textbf{962} \\ 

\cline{2-9}
& \rankS  & .798 & .714 & .789 & .796 & .717 & .899 & .967 \\

\hline
\hline
\multirow{4}{*}{\rotatebox{90}{\rev{k=3}}}   

& \sBert & .842 & .767 & .832 & .838 & .773 & .925 & .999 \\
&  \rwWord  & .841 & .766 & .832 & .836 & .775 & .926 & .974 \\
&  \rwWordEx  & .\textbf{858} & .\textbf{787} & .\textbf{848} & .\textbf{854} & .\textbf{795} & .\textbf{887} & .988    \\   
\cline{2-9}
& \rankS  & .890 & .830 & .887 & .890 & .830 & .969 & .994 \\

\end{tabular}
\end{table}

\vspace{1ex}
\noindent \textbf{Matching results.}
\rev{Results in Tables~\ref{tab:politi}, \ref{tab:snopes} and \ref{tab:sts} 
show that our method is the best unsupervised solution, outperforming \sBert in all measures and scenarios. 
Table~\ref{tab:sts} shows that as sentences with higher similarities share more details, increasing $k$ improves the performance for all methods.} Our approach sits between the best unsupervised baseline and the supervised method. One explanation is that these datasets contain generic textual claims with common terms, which is the best scenario for pre-trained models trained on very large corpora. Also, long natural language English sentences are nicely modelled by the attention mechanism in transformers. As we discuss in Section~\ref{subsec:improve}, by combining our embeddings with pre-trained language models, we can improve our performance.

\begin{table}[t]
	\centering
	\footnotesize
	\caption{\rev{Train and test execution times (sec).}}
	\vspace{-1ex}
	\label{tab:exetime}
		\begin{tabular}
        {c|c c|c c|c c}
        {\it Method} & \multicolumn{2}{c|}{Text to data} &
        \multicolumn{2}{c|}{Structured text} &
        \multicolumn{2}{c}{Text to text} 
        \tabularnewline
		&
		{\it Train}&{\it Test}&
		{\it Train}&{\it Test}&
		{\it Train}&{\it Test} \tabularnewline
		\hline
			\wvec  & 13.9 & 239 & 3.5 & 11.82 & 5.0 & 107  \tabularnewline 
			\hline
			\dvec  & 47.7 & 17.2 &  8.5 &  1.30 & 14.9 & 21.95 \tabularnewline
			\hline
			\sBert  & - & 2.6 &  - &  1.16 & - & 7.5
			\tabularnewline
			\hline
			\rwWord  & 152 & 0.07 & 207 & 0.05 & 189 & 0.41 \tabularnewline 
			\hline
			\hline
			\hline
			\rankS  & 3206 & 0.09 &  3916  &  0.05 & 6918 & 1.2  \tabularnewline
			\hline
			\lBertS   & 3616 & 0.25 & 3280 &  0.69 & 251 & 2.6 \tabularnewline
			\hline
			\rev{\tapas}  & 7301 & 7.2 & -  &  - & - & - 
			\tabularnewline
			\hline
			\deep  & 3492 & 0.68 & -  &  - & - & - 
			\tabularnewline
			\hline
			\ditto  & 34528 & 2.28 & -  &  - & - & - 
			\tabularnewline
		\end{tabular}
	\vspace{-1ex}
\end{table}

\subsection{Compression Results}
We report the performance of the compressing technique introduced in Section~\ref{sec:expand}. As a baseline technique, we report also for \textit{SSuM}, a state of the art method that employs node merging and edge sparsifying to generate a super-graph as output~\cite{lee2020ssumm}.
 Table~\ref{tab:compression} compares the performance of compression methods in terms of \textit{size} (number of nodes and edges) of the compressed graph and of \textit{quality} in the matching task (MRR).

\begin{table*}[t]
\small
\centering
\begin{tabular}{c|ccc||ccc||ccc|ccc|ccc}
 Dataset & \multicolumn{3}{c}{Original Graph} & \multicolumn{3}{c}{Expanded Graph} & \multicolumn{3}{c}{MSP (0.5)} & \multicolumn{3}{c}{MSP (0.25)} & \multicolumn{3}{c}{SSuM (0.1)} \\
 & \#N & \#E & MRR & \#N & \#E & MRR & \#N & \#E & MRR & \#N & \#E & MRR & \#N & \#E & MRR \\
 \hline
IMDB & 107k & 1m & .780 & 237k & 1.5m & \textbf{.792} & 82k & 887k & .779 & 75k & 840k & .755  & \textbf{27k} & \textbf{540k} & .601 \\
Corona & 10k & 43k & .728 & 15k & 56k & .755 & 10k & 40k & \textbf{.769} & \textbf{8.5k} & \textbf{32.5k} & .757 & 10k & 33k & .610 \\
Snopes & {35k} & {129k} & .695 & 142k & 622k & \textbf{.708} & 84k & 479k & .647 & \textbf{49k} & \textbf{292k} & .586 & 83k & 470k & .590 \\
Politi & {24k} & {168k} & {.489} & 62k & 317k & \textbf{.507} & 37k & 242k & .500 & {\textbf{33k}} & {\textbf{225k}} & .484 & 52k & 257k & .397 \\
Audit & 6k & 164k & {.421} & 17k & 202k & \textbf{.452} & 7k & 161k & .389 & \textbf{5.5k} & \textbf{144k} & .362 & 14k & 150k & .392
\end{tabular}
\caption{Compression performance: number of graph nodes (\#N) and edges (\#E) compared with matching quality MRR.}
\label{tab:compression}
\end{table*}

For \textit{MSP}, we report results for iterations equal to half (\textit{MSP (0.5)}) and a quarter (\textit{MSP (0.25)}) of the expanded graph's nodes. \textit{SSuM (0.1)} is set with a compression ratio of $0.9$ as this is the value generating the best quality results (MRR) in our experiments. 

In terms of size reduction, \textit{MSP (0.25)} is the compression method with the best results in four cases and it is second to \textit{SSuM} only for IMDB. However, it shows an higher decrease in match quality results w.r.t. \textit{MSP (0.5)}, which is the compression method with the best results in all cases. For Corona it even does better than the expanded graph. This dataset contains many numerical values (about 25\% of its data nodes in the expanded graph), which are misleading in some cases, as they are more likely to raise spurious connections in the graph. 
In general, \textit{MSP} performs better than \textit{SSuM}. For higher compression, \textit{MSP (0.25)} produces smaller graphs with better match accuracy in most cases. For \textit{MSP (0.5)}, we observe better quality in the matches in all cases except Audit, and comparable size in the compressed graphs. The results show the benefit of considering shortest paths among metadata nodes. For \textit{MSP}, the  graph size and the match accuracy follow the expected behavior w.r.t. the compression ratio.

\textit{MSP}, in both executions, shows its best results for scenarios with at least one relational table. In these cases, the compressed graph is smaller than the original one (and much smaller than the expanded one), with better or very close matching quality. For text-only scenarios, the size reduction is remarkable, and better than \textit{SSuM}, but a significant drop in matching quality can be observed. The conclusion is that the use graph compression depends on the kind of data and the requirements for the target application at hand.

\subsection{Execution Times.}
Table~\ref{tab:exetime} reports execution times for all methods averaged over the experiments for every task. For training time, embeddings methods (\wvec, \dvec) and transformer-based methods (\rankS, 
\lBertS) are trained (fine tuned) on a smaller corpus than our method (\rwWord). This is because we create 100 walks for each node in the graph, which leads to bigger corpora in general. Due to this difference, \textit{Word2Vec} and \textit{Doc2Vec} are faster than other methods in training, while our method has execution times smaller than those taken to fine tune transformers. \sBert has no training.

We report the average execution time for a single match (test). Our solution is the fastest. Document based methods, like \textit{Doc2Vec}, are faster than word based embedding solutions. This is because for the latter methods we generate vectors for \textit{all} tokens in the document and aggregate them. Classifiers are faster than document embeddings based matching, but slower than our method. In the training step of our method, expansion and compression take less than 3k seconds in all cases, with the exception of IMDB (by far the largest) with 79k seconds for expansion (with DBpedia) and 51k seconds for compression with \textit{MSP (0.5)}.

\subsection{Ablation Study} \label{sec:ablation}
We first report on the impact of different parameters on the performance of our method \rwWord. We then evaluate the impact of the improvements proposed in Section~\ref{sec:method}.

\subsubsection{Impact of parameters} \label{subsec:impacts}
We examine the impact of length and number of random walks, followed by number of tokens in data nodes. \rev{We also report the impact of the graph size (in terms of number of nodes) on the execution time of our model.}
For \textit{CoronaCheck}, we report results for the union of the \textit{Generated} and \textit{User} sentences.

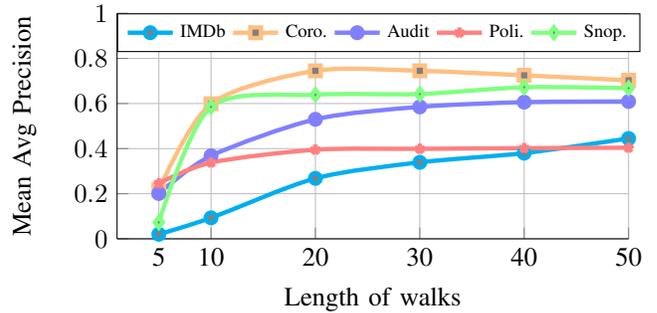
\begin{figure}
\begin{tikzpicture}
\begin{axis}[
    width=6.8cm,
   height=3cm,
   scale only axis,
   xmin=1, xmax=50,
   xtick={5,10,20,30,40,50},
   xticklabels={5,10,20,30,40,50},
   xmajorgrids,
    xlabel= Length of walks,
    ylabel= Mean Avg Precision,
   ymin=0.0, ymax=1.0,
   ymajorgrids,
   axis lines*=left,
   legend style ={ at={(0,1.0)},legend columns=5, font = \scriptsize,
        anchor=north west,  
        fill=white,align=left},
    cycle list name=black white,
    smooth
] 
    \addplot+[color=cyan,line width=1.64pt]
    coordinates{(5,0.020)(10,0.093)(20,0.268) (30,0.339) (40,0.380) (50,0.445) };
    \addlegendentry{IMDb};

    \addplot+[color=orange!50,line width=1.64pt]
    coordinates{(5,0.216)(10,0.599)(20,0.745) (30,0.745) (40,0.725) (50,0.703) };
    \addlegendentry{Coro.};
    
    \addplot+[color=blue!50,line width=1.64pt]
    coordinates{(5,0.201)(10,0.368)(20,0.530) (30,0.585) (40,0.606) (50,0.609) };
   \addlegendentry{Audit};

    \addplot+[color=red!50,line width=1.64pt]
    coordinates{(5,0.246)(10,0.338)(20,0.395) (30,0.399) (40,0.402) (50,0.404) };
   \addlegendentry{Poli.};

    \addplot+[color=green!50,line width=1.64pt]
    coordinates{(5,0.073)(10,0.585)(20,0.639) (30,0.642) (40,0.672) (50,0.668) };
   \addlegendentry{Snop.};

   \end{axis}
\end{tikzpicture}%
\vspace{-1ex}
  \caption{Match quality with increasing walk length.}
  \label{fig:lengh_walks}
\vspace{-2ex}
\end{figure}

\vspace{1ex}
\noindent \textbf{Length of random walks.} 
Figure~\ref{fig:lengh_walks} shows the mean average precision results for all scenarios when increasing the length of the random walks. 
Increasing the walk length increases the performance for all scenarios up to size 20. The  increase is higher at lower values and then stabilizes or gradually decreases for most scenarios. 
We explain the different behavior for \textit{IMDb} and \textit{Audit}
with the fact that they have the biggest and most dense graphs. \textit{IMDb} graph is the biggest both in terms of nodes (107k nodes vs 35k node for \textit{Snopes}) and edges (1m vs 168k edges for \textit{Politifact}). 
Because of their size, larger graphs benefit of walks longer than 20.

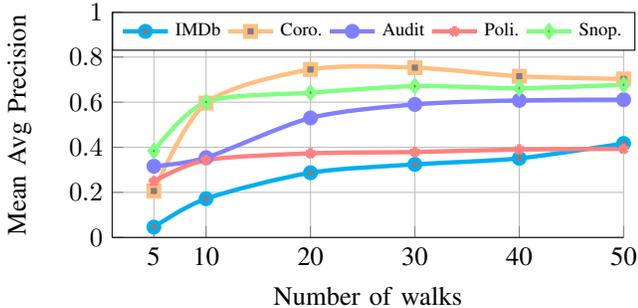
\begin{figure}[t]
\begin{tikzpicture}
\begin{axis}[
    width=6.8cm,
   height=3cm,
  scale only axis,
  xmin=1, xmax=50,
  xtick={5,10,20,30,40,50},
  xticklabels={5,10,20,30,40,50},
  xmajorgrids,
    xlabel= Number of walks,
    ylabel= Mean Avg Precision,
  ymin=0.0, ymax=1.0,
  ymajorgrids,
  axis lines*=left,
  legend style ={ at={(0,1.0)},legend columns=5, font = \scriptsize,
        anchor=north west,  
        fill=white,align=left},
    cycle list name=black white,
    smooth
]

\addplot+[color=cyan,line width=1.64pt]
    coordinates{(5,0.046)(10,0.172)(20,0.287) (30,0.324) (40,0.351) (50,0.417) };
    \addlegendentry{IMDb};

    \addplot+[color=orange!50,line width=1.64pt]
    coordinates{(5,0.206)(10,0.596)(20,0.745) (30,0.753) (40,0.715) (50,0.703) };
    \addlegendentry{Coro.};
    
    \addplot+[color=blue!50,line width=1.64pt]
    coordinates{(5,0.316)(10,0.354)(20,0.530) (30,0.590) (40,0.608) (50,0.611) };
  \addlegendentry{Audit};

    \addplot+[color=red!50,line width=1.64pt]
    coordinates{(5,0.249)(10,0.343)(20,0.373) (30,0.379) (40,0.390) (50,0.393) };
  \addlegendentry{Poli.};

    \addplot+[color=green!50,line width=1.64pt]
    coordinates{(5,0.384)(10,0.601)(20,0.642) (30,0.671) (40,0.662) (50,0.677) };
  \addlegendentry{Snop.};

  \end{axis}
\end{tikzpicture}%
\vspace{-1ex}
 \caption{Increasing number of random walks per node.}
 \label{fig:number_walks}
\vspace{-2ex}
\end{figure}

\noindent \textbf{Number of walks.} 
Figure~\ref{fig:number_walks} shows the mean average precision when increasing the number of walks.  The performance for all datasets improve with more walks, but with diminishing results. Results also confirm that graphs with more edges per node need more walks to obtain the best results. After 20 walks per node, results for \textit{IMDb} keep improving, while for \textit{CoronaCheck}, which is the most sparse graph with an average of 
four edges per node, there is no improvement.

\noindent \textbf{Number of tokens in terms.} 
Allowing more tokens in data nodes (terms) increases the mean average precision in all scenarios. 
There is a significant increase in quality going from one to two tokens and the impact is smaller with higher values. The highest increase is for \textit{Snopes} and \textit{Audit} datasets with an increase up to $0.24$ and $0.07$, respectively. For text to data scenarios, \textit{IMDb} has an increase up to $0.05$ and \textit{CoronaCheck} up to $0.03$. The amount of increase for a scenario is related to the number of new nodes added to the graph. For \textit{Snopes}, an increase in the number of tokens in a term adds an average of 24k new nodes to the graph, which is close to the initial number of nodes (35K). For \textit{Audit}, the increase is 1.7k, which is $62\%$ of the initial graph's token nodes. In \textit{IMDb}, 23K new nodes are generated ($22\%$ of initial graph) at every increase. 
For all scenarios, except \textit{Snopes}, the number of nodes does not increase drastically after allowing three tokens in a term. 
The behaviour of a dataset is determined by the length of documents in its first corpus as tokens in the documents of the second corpus get filtered. \textit{Snopes} has the biggest documents in its first corpus in comparison to the other datasets: it has claims of 43 tokens on average while \textit{PolitiFact} has 18.

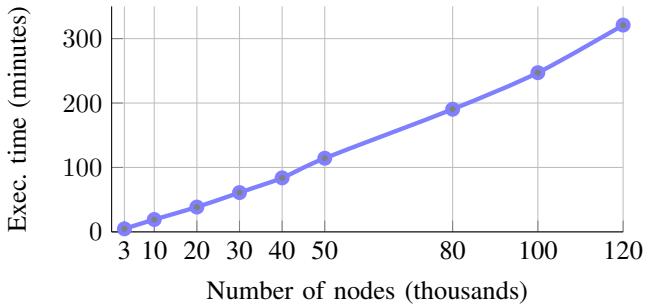
\begin{figure}[t]
\begin{tikzpicture}
\begin{axis}[
    width=6.8cm,
   height=3cm,
  scale only axis,
  xmin=0, xmax=120,
  xtick={3,10,20,30,40,50,80,100, 120},
  xticklabels={3,10,20,30,40,50,80,100,120},
  xmajorgrids,
    xlabel= Number of nodes (thousands),
    ylabel= Exec. time (minutes),
  ymin=0, ymax=350,
  ymajorgrids,
  axis lines*=left,
    cycle list name=black white,
    smooth
]

    \addplot+[color=blue!50,line width=1.64pt]
    coordinates{(3,4.7)(10,19.2)(20,38.5) (30,61) (40,83.8) (50,114.3) (80,190.4) (100,247.1) (120,321) };

  \end{axis}
\end{tikzpicture}%
\vspace{-1ex}
 \caption{\rev{Increasing number of nodes in the graph.}}
 \label{fig:number_nodes}
\vspace{-2ex}
\end{figure}

\rev{
\noindent \textbf{Number of nodes in the graph.} 
To show that our method can scale on a commodity machine, we generate graphs of increasing size. The graphs are generated from the \textit{STS} dataset with \textit{k} values between 0 and 5. For increasing further the size of the graphs, we expand them using Algorithm~\ref{alg:graphExpansion} and \textit{ConceptNet}.  
For every graph, we generate 100 random walks of length 30 for each node and report in Figure~\ref{fig:number_nodes} the total time to generate random walks and to train the word embeddings. Results show that by increasing the number of nodes in the graph, the execution time increases linearly.}

\subsubsection{Improving graph generation} \label{subsec:improve}
We discuss here the impact of the techniques introduced in Section~\ref{sec:method} to improve embeddings and matching quality. 

\vspace{1ex}
\noindent \textbf{Connecting metadata nodes.} 

For measuring the impact of edges between related metadata nodes in a structured text, we run the same experiments in Section~\ref{sec:auditExp} for \rwWord without edges among metadata nodes. The quality of the matches is negatively affected, with the Node F-scores
for increasing values of $K$ (1, 3, 5, 10) 
dropping by $.08$, $0.04$, $0.02$, and $0.01$ in absolute values. 

\vspace{1ex}
\noindent \textbf{Filtering data nodes.} 
Merging tokens create shorter paths between related metadata nodes. In graph creation, we filter tokens of the second corpus based on the nodes from the first one. In this experiment, we compare the performance of our technique (Intersect) against a solution based on \textit{TF-IDF}~\cite{DITTO21}, 
which retains non-stopword tokens with high
\textit{TF-IDF} scores. It has been shown that in a data to data matching task this technique  increases the quality performance on a text-heavy dataset from 41$\%$ to $93\%$~\cite{DITTO21}.
For implementing this method, for each document we keep $k$ tokens with highest TF-IDF scores for different $k$ values. For each scenario, we run $k=$3, 5, 10 and 20 and report the best result. 

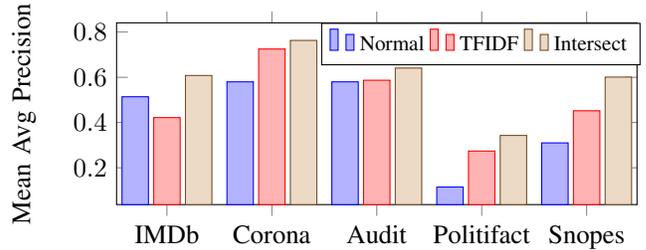
\begin{figure}[t]
        \centering
\begin{tikzpicture}
\begin{axis}[
    ybar,
   width=8.5cm,
   height=4cm,
    enlargelimits=0.12,
    legend style={at={(0.7,1)},
      anchor=north,legend columns=-1, font = \footnotesize},
    ylabel={Mean Avg Precision},
    symbolic x coords={IMDb,Corona,Audit,Politifact,Snopes},
    xtick=data,
    nodes near coords align={vertical},
    ]
\addplot coordinates {(Audit,0.580) (Politifact,0.115) (Snopes,0.310) (IMDb,0.514) (Corona,0.580)};
\addplot coordinates {(Audit,0.587) (Politifact,0.274) (Snopes,0.452) (IMDb,0.422) (Corona,0.725)};
\addplot coordinates {(Audit,0.641) (Politifact,0.343) (Snopes,0.601) (IMDb,0.608) (Corona,0.763)};

\legend{Normal,TFIDF,Intersect}
\end{axis}
\end{tikzpicture}
    \caption{Impact of data node filtering.}
    \label{fig:summarizing}
\vspace{-2ex}
\end{figure}

Figure~\ref{fig:summarizing} shows the performance of these two techniques in terms of mean average precision for all scenarios. The results show that, except for \textit{IMDb} dataset and \textit{TF-IDF},
both summarizing techniques improve the mean average precision of matching. It also shows that our technique works better than \textit{TF-IDF} in all scenarios.

\begin{figure}
    \centering
    \begin{tikzpicture}
    \begin{axis}[
        ybar,
            width=8cm,
   height=4cm,
    enlargelimits=0.15,
        legend style={at={(0.63,1.03)},
          anchor=north,legend columns=-1,font = \footnotesize},
        ylabel={Mean Avg Precision},
        symbolic x coords={IMDb,Corona,Audit,Politifact,Snopes},
        xtick=data,
        nodes near coords align={vertical},
        ]
    \addplot coordinates {(IMDb,0.598) (Corona,0.725) (Audit,0.626) (Politifact,0.401) (Snopes,0.693)};
    
    \addplot coordinates {(IMDb,0.603) (Corona,0.803) (Audit,0.646) (Politifact,0.445) (Snopes,0.789)};
    \legend{\rwWord,\rwWord\&\sBert}
    \end{axis}
    \end{tikzpicture}
    \vspace{-1ex}
    \caption{Our method combined with \textit{SentenceBERT}.}
    \label{fig:ensemble}
    \vspace{-2.5ex}
\end{figure}
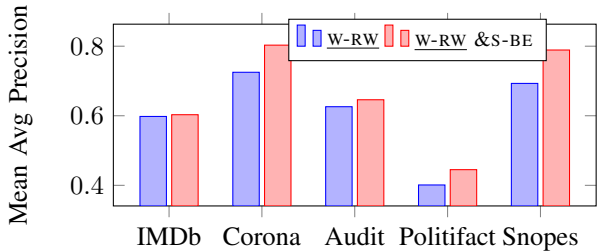

\vspace{1ex}
\noindent \textbf{Combining matching scores.} 
Figure~\ref{fig:ensemble} depicts the results of averaging the cosine similarity scores from our solution with those from the pre-trained \sBert.
Our solution already outperforms \sBert in the original experiments, but averaging the two methods improves the matching quality in all scenarios even above the results from our methods alone. 
The biggest improvements are for \textit{Snopes} and \textit{CoronaCheck} with $0.9$ and $0.8$ increase, respectively. This simple combination show the benefit of exploiting domain specific embeddings and pre-trained ones from large, generic corpora.

\vspace{1ex}
\noindent \textbf{Merging nodes.} 
For all scenarios, we employ different techniques and resources to improve their quality results (Section~\ref{subsec:external}). In \textit{CoronaCheck}, $17\%$ of the nodes are numeric values and merging with equal-width buckets decreases the number of graph nodes. We had the best results with equal-width buckets of size 7, which increased the mean average precision from $0.72$ to $0.76$. In \textit{Audit}, \textit{Politifact}, and \textit{Snopes} datasets, there are few numeric values and bucketing had no effects. In \textit{IMDB}, we observe a small loss because numeric values are release dates, for which is better to avoid merging.

We also use \textit{Wikipedia2Vec} to merge similar nodes. \textit{IMDb} contains variations for the same entity (e.g., director names) and merging them with a threshold $\gamma=0.57$ increases the performance by $2.5\%$. For \textit{Snopes} and \textit{Politifact}, using the same value, the increase is by $1.7\%$ and $1.5\%$, respectively. As \textit{CoronaCheck} contains typos in user sentences (e.g., country names), merging such typos 
leads to a $3.4\%$ increase.  
\textit{Auditing} does not show improvements by merging nodes with pre-trained resources because of the difference between the general meaning of a term and its meaning in a specific domain. Models pre-trained on general corpora do not help much in a domain specific scenario. For example, the \textit{Wikipedia2Vec} similarity is high between \textit{Financial statement} and \textit{Financial reporting}, same for \textit{Auditor} and \textit{Risk control}. However, these terms have different meanings in audit documents.

\section{Related Work} \label{sec:related}
Entity resolution for relational data~\cite{getoor2013entity} has been recently studied with deep learning solutions~\cite{CappuzzoPT20,EbraheemTJOT18,DITTO21}. As they rely on the presence of a schema, one way to use them in this setting is to treat the text paragraphs as tuples within a single column, but this leads to poor results. 
\rev{Even approaches that consider schema-agnostic attribute matching, assume a semi-structured input~\cite{Mandilaras0GSTG21,SinghMEMPQST17}. In contrast, }
our solution models texts and tables in a unified representation for learning embeddings, with benefits in the matching tasks even with very lightweight algorithms.
\rev{Compared to~\cite{CappuzzoPT20}, we extended the graph generation to handle free text (without schema) with new merging and tokenization techniques and with a new expansion and compression approach.}
We model text matching 
both as a binary classification task and as a multi-label classification task to use SOTA baselines based on fine tuning language models~\cite{ShaarBMN20,adhikari2019docbert}. These methods outperform traditional IR approaches, such as BM25,
but do not focus on the problem of matching text and relational datasets. This latter problem has been studied in settings that do not cover our use cases, either because they assume supervision or because they are no domain specific~\cite{IbrahimRWZ19,KaragiannisSPT20,ChenWCZWLZW20}. Other approaches for text and data matching assume a very expensive training over large document corpora and millions of tables~\cite{HerzigNMPE20,YinNYR20,DengSL0020}. We \rev{report results for the fine-tuning on top of these pre-trained model (\tapas), but do not report results based on pre-training as they are significantly worse in terms of quality w.r.t. our solution. Indeed, the pre-training is not designed for a setting with a small table corpus}. 
Our setting is also different from the problem of entity linking, as we are matching long text to tuples in relational data~\cite{chen2019yelm}, \rev{but it could be used as a pre-processing tool to provide correspondences to knowledge translation solutions~\cite{BashardoostMLN20}.}

Our default method to generate data and metadata representations is \emph{Word2Vec}~\cite{mikolov2013efficient} 
as it is powerful in discovering relationships in the corpus as well as similarity between tokens~\cite{kowsari2019text}. While we generate embeddings, some of our baselines use pre-trained ones
~\cite{pennington2014glove,devlin2018bert}.
Document embedding methods model longer text sequences by 
aggregating the vectors of words in the given sentence or paragraph~\cite{kiros2015skip,de2016representation}
or by learning the document vector with special tokens~\cite{
adhikari2019docbert,reimers2019sentence}.
Given our graph, it is also possible to generate embeddings directly for its nodes~\cite{goyal2018graph,perozzi2014deepwalk,grover2016node2vec}
with comparable results w.r.t. the random walks followed by word embedding generation in the data to data matching task~\cite{CappuzzoPT20}. Our study confirms that they do not bring clear benefit, but are more resources intensive than \textit{Word2Vec}. Finally, our work can be seen as a new instance of the general approaches of using deep learning for data integration~\cite{FuHSCZWK19,FernandezM19,abs-2010-07386} and of improving pre-trained embeddings w.r.t. relational data~\cite{0002OTL20}.

\section{Conclusion}
\label{sec:conclusion}
We presented a 
new generic matching task that allows both structured text documents and relational data. Results show that lightweight embeddings 
effectively model the similarity between heterogeneous corpora. Our proposal outperforms in quality and test time all unsupervised baselines and it is competitive to supervised solutions.
Our graph expansion always leads to the best matching quality, while the compression is effective in reducing the graph size, but comes with a trade off in the performance of the matching for text-only corpora. We plan to extend our framework to support a richer graph with typed edges and 
blocking to speed up performance.
\rev{However, while word embeddings are effective and efficient to generate, a more complex architecture should be considered to enable the benefits of fine tuning when examples are available.}

\bibliographystyle{IEEEtran}
\bibliography{DA.bib}
\end{document}